# Calculation of Forces to the High Granularity Calorimeter Stainless Steel Absorbers Misaligned inside the CMS Superconducting Coil

Vyacheslav Klyukhin[1,2] for the CMS Collaboration

[1*]Skobeltsyn Institute of Nuclear Physics, Lomonosov Moscow State University, RU-119991, Moscow, Russia
[2]CERN, CH-1211, Geneva 23, Switzerland

[*]Corresponding author. E-mail : Vyacheslav.Klyukhin@cern.ch; ORCID: 0000-0002-8577-6531

**Abstract**

The general-purpose Compact Muon Solenoid (CMS) detector at the Large Hadron Collider (LHC) at CERN includes the hadronic calorimeter to register the energies of the charged and neutral hadrons produced in the proton-proton collisions at the LHC at a centre of mass energy 13.6 TeV. The calorimeter is located inside the superconducting solenoid of 6 m in diameter and 12.5 m in length operating with a direct current of 18.164 kA that creates the central magnetic flux density of 3.81 T. To fit optimally to the high pileup and high radiation environment of the High Luminosity LHC, the existing CMS endcap calorimeter will be replaced with a new high granularity calorimeter (HGCal) comprising the electromagnetic and hadronic sections. The hadronic section in each of two endcaps of the HGCal will include 22 stainless steel absorber plates with a relative permeability value of 1.000588 at to the central magnetic flux density of 3.81 T. The calculation of the electromagnetic forces to the absorber plates due to orthogonal and angular misalignments of the HGCal endcaps with respect to the coil axis is performed with a three-dimensional computer model of the CMS magnet using the method of calculation described earlier.

## 1 Introduction

The Compact Muon Solenoid (CMS) multi-purpose detector [1] at the Large Hardon Collider (LHC) registers the charged and neutral particles created in the proton-proton collisions at a center of mass energy 13.6 TeV. The detector includes a wide-aperture superconducting thin solenoid coil with a diameter of 6 m, a length of 12.5 m, and a central magnetic flux density $B_0$ of 3.81 T created by the operational direct current of 18.164 kA. Inside the superconducting coil around the interaction point of proton beams the major particle subdetectors are located: a silicon pixel and strip tracking detectors to register the charged particles; a solid crystal electromagnetic calorimeter to register electrons, positrons and gamma rays; a barrel and endcap hadronic calorimeters of total absorption to register the energy of all the hadron particles. Outside the solenoid coil the muon spectrometer chambers register muon particles escaping the calorimeters.

In the next run of the LHC, the High Luminosity (HL) operational phase is scheduled after a long shutdown of the machine. For optimal operation in the high pileup and high radiation environment of the HL-LHC the existing CMS plastic scintillator-based hadron endcap calorimeter will be replaced by a new High Granularity Calorimeter (HGCal) containing the silicon sensors and plastic scintillators as active material and stainless-steel absorber plates in the hadronic compartments of two endcaps [2]. The slightly magnetic stainless steel plates occupy at each endcap a volume about 21 m$^3$. Assuming to be produced with a relative permeability $\mu_{rel}$ well below 1.05 they nevertheless attract in the magnetic field of 3.81 T to the center of the CMS superconducting coil with electromagnetic axial forces of order of 19 kN assuming $\mu_{rel}$ value to be 1.000588 at this magnetic field [3]. In the present study, with a three-dimensional (3D) CMS magnet model [4] based on a 3D finite-element code TOSCA (two scalar potential method) [5], developed in 1979 [6] at the Rutherford Appleton Laboratory, the forces on each HGCal hadronic compartment are calculated for $\mu_{rel}$ equal to 1.000588 at the $B_0$ of 3.81 T for small misalignments of the HGCal endcaps with respect to the CMS coil axis.

The article is organized as follows: Section 2 describes the model of the HGCal stainless steel absorber plates inside the CMS superconducting coil; Section 3 contains results of the magnetic force calculations to each entire endcap hadronic compartment for two types of misalignments: small displacements and rotations; and, finally, conclusions are drawn in Section 3.

## 2 Basic HGCal hadronic compartment model

In Figure 1 a perspective view of the CMS detector is displayed with existing electromagnetic and hadronic endcap calorimeters colored with green and yellow colors, respectively. The calorimeters are shown at one side of the solenoid coil. At another side their positions are symmetrical with respect to the detector middle plane. For the HL-LHC operations both endcap calorimeters will be replaced with the HGCal endcaps.

In this study we use the reference CMS coordinate system where the origin is in the center of the superconducting solenoid, the $X$ axis lies in the LHC plane and is directed to the center of the LHC machine, the $Y$ axis is directed upward and is perpendicular to the LHC plane, the $Z$ axis makes up the right triplet with the $X$ and $Y$ axes and is directed along the vector of magnetic flux density $\boldsymbol{B}$ created on the axis of the superconducting coil.

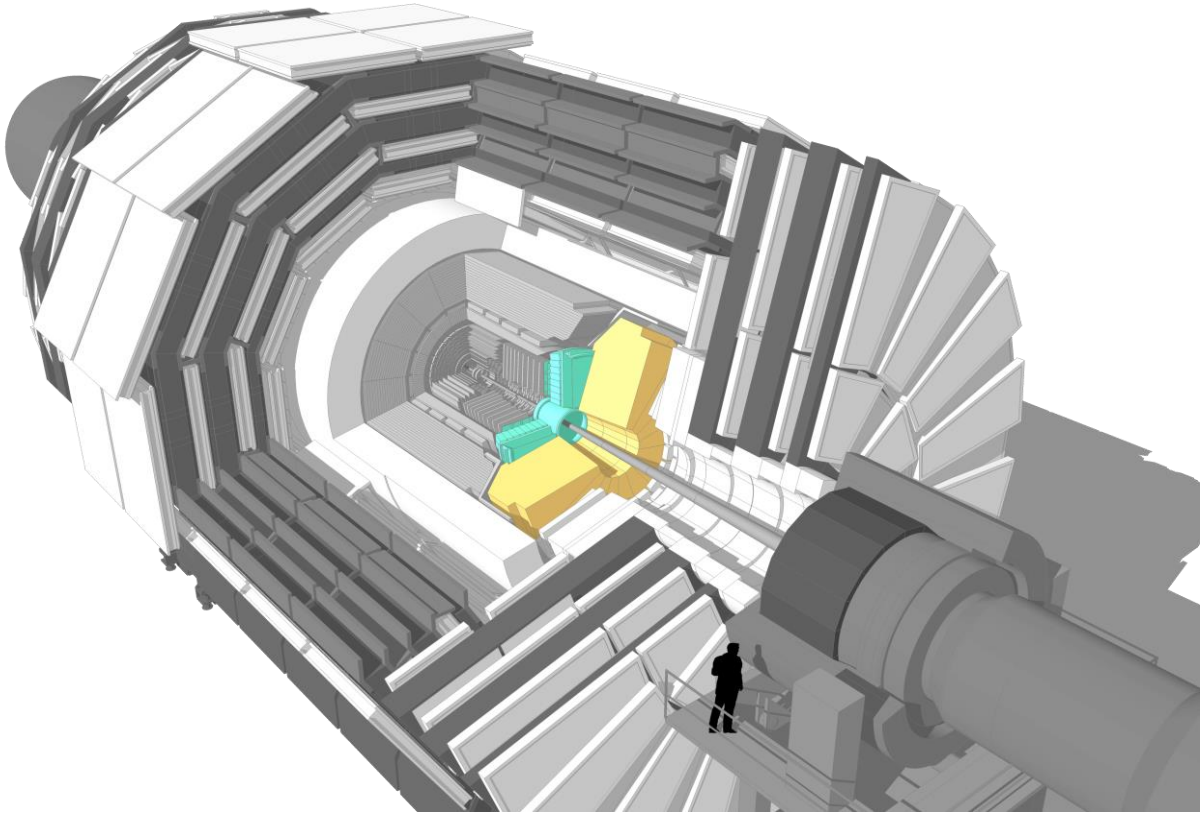

**Figure 1.** Perspective view of the CMS detector with 1/8 part opened. Inside this part an existing electromagnetic endcap calorimeter is shown with green color, and an existing hadronic endcap calorimeter is shown with yellow color. Positions of both endcap calorimeters are symmetrical with respect of the detector middle plane, and both calorimeters will be replaced with the HGCal.

To absorb the hadronic particles along the entire length of each HGCal endcap, 22 stainless steel disks with sicknesses from 45 to 95.4 mm interleaved with the air gaps of 21.55 mm form in each endcap the shapes shown in Figure 2 at the positive $Z$ coordinate values, and Figure 3 at the negative $Z$ coordinate values. In the CMS magnet model, the disks on both sides of the HGCal are located at the distances from 3.6098 to 5.22391 m from the coil middle plane. These distances include the displacements by 12 mm directed to the coil center on both sides according to the magnet yoke deformation under the magnetic forces when the magnet is switched on. The disks are numerated from 1 to 22 from the smallest disk to a back disk in positive $Z$ direction, and from −1 to −22 in the same way in negative $Z$ direction.

Figures 2 and 3 are prepared with the updated latest CMS magnet 3D model [3,4]. The model is calculated with the operational direct current of 18.164 kA, and with the stainless steel $\mu_{rel}$ value of 1.000588 that corresponds to central magnetic flux density in the CMS coil of 3.81 T. As shown in Figures 2 and 3, the absorber disks form two cones (small and large) and two cylinders at each side of the coil inner volume. The smallest outer diameter of the small cone is 3.2934 m, the outer diameter of the large cylinder is 5.2492 m, and the volume of the absorber plates modelled in each HGCal endcap is 18.9 m$^3$. The small cone has an angle of 19.09° with respect to the coil axis, and a similar angle of the large cone is 52.67°. An angle in transition disks with numbers ±5 between the small and large cones is 44.93°. In each endcap the first eight disks in the conical part (with numbers from ±1 to ±8) have an inner bore with a radius of 0.3136 m; the next four disks (with numbers from ±9 to ±12) have an inner bore with a radius of 0.3836 m; the last two disks in the conical part (with numbers ±13 and ±14) and all the eight cylinder disks (with numbers from ±15 to ±22) have an inner bore of 0.4448 m.

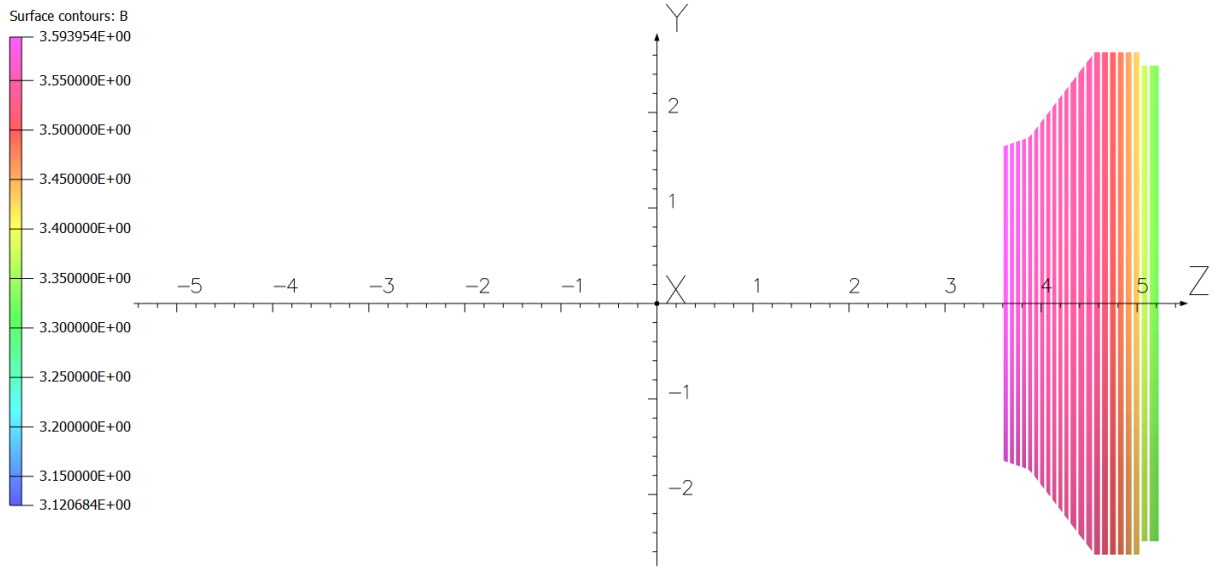

**Figure 2.** HGCal stainless steel absorber plates in the longitudinal section of the CMS coil inner volume at positive $Z$ coordinates. The color scale of the total magnetic flux density $B$ on the disk surfaces is in Tesla with a unit of 0.05 T. The maximum value of $B$ is 3.594 T, the minimum value is 3.121 T. The values on the coordinate axes are given in meters. The plates are numbering from 1 to 22 from origin of the coordinate system to positive $Z$ values. This "positive" endcap compartment contains 22 stainless steel disk plates with thicknesses of 45 (disk number 1), 41.5 (disk numbers from 2 to 11), 60.7 (disk numbers from 12 to 21), and 95.4 (disk number 22) mm. The air gaps between the disks are of 21.55 mm.

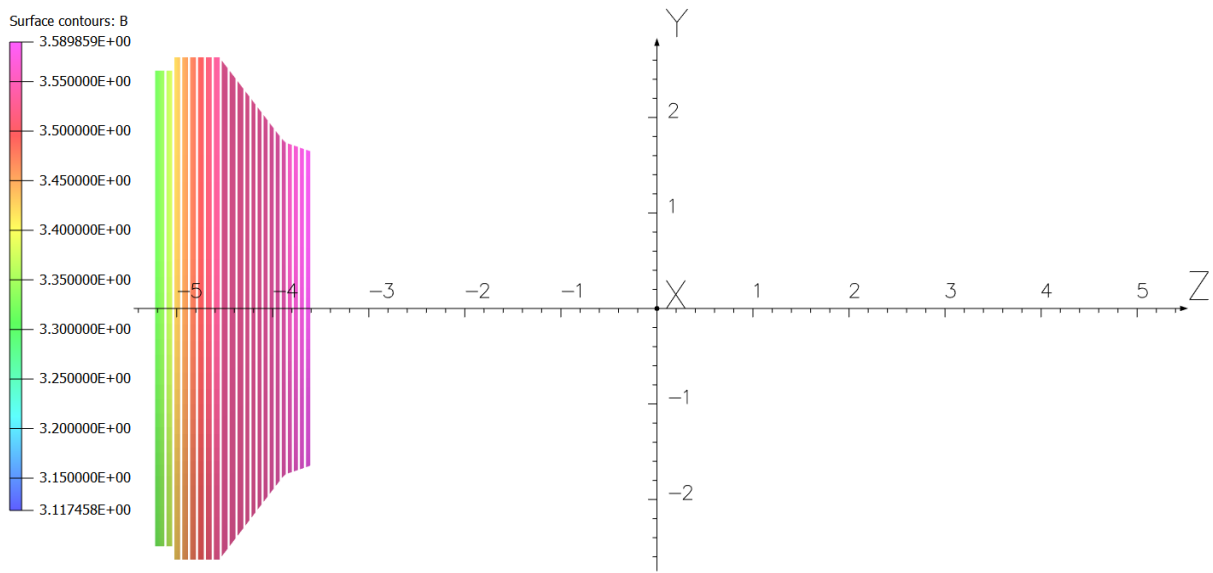

**Figure 3.** HGCal stainless steel absorber plates in the longitudinal section of the CMS coil inner volume at negative $Z$ coordinates. The color scale of the total magnetic flux density $B$ on the disk surfaces is in Tesla with a unit of 0.05 T. The maximum value of $B$ is 3.590 T, the minimum value is 3.117 T. The values on the coordinate axes are given in meters. The plates are numbering from −1 to −22 from origin of the coordinate system to negative $Z$ values. This "negative" endcap compartment contains 22 stainless steel disk plates with thicknesses of 45 (disk number −1), 41.5 (disk numbers from −2 to −11), 60.7 (disk numbers from −12 to −21), and 95.4 (disk number −22) mm. The air gaps between the disks are of 21.55 mm.

In Figures 2 and 3, a distribution of the magnetic flux density total component $B$ on the absorber plate surfaces is shown. For both "positive" and "negative" endcaps $B$ have a maximum value about 3.59 T, and a minimum value about 3.12 T. The difference in $B$ on "positive" and "negative" endcaps arises from one missing turn in the inner layer of conductor in the coil module at the negative $Z$ coordinate of −3.763765 m. The radial component of the magnetic flux density $B_r$ has an extremum value of ±0.41 T on the cylinder surfaces of the 60.7 mm thick disks with numbers ±20.

The central magnetic flux density $B_0$ is equal to 3.8095 T and is 0.00079 % greater than that value in the present CMS configuration without the stainless steel absorber plates. Thus, with $\mu_{rel}$ value of 1.000588, a contribution of the stainless steel plate magnetization into the CMS inner field is extremely small.

The force $F$ acting to the stainless steel absorber plate with surface $S$ located in the magnetic field with the magnetic flux density $B$ is calculated by integrating the Maxwell stress tensor [7] over the absorber surface. The following formula is valid for the integration in the air near the absorber surface:

$$F = \frac{1}{\mu_0} \int_S \left[ B (B \cdot n) - \frac{B^2}{2} n \right] dS, \tag{1}$$

where $\mu_0$ is a permeability of free space and $n$ is a unit vector of the outer normal to the integration surface $S$. The force calculation procedure [3] includes calculating the axial forces using Eq. (1) in the model where $\mu_{rel}$ in the absorber plates is equal to 1.000588, and a subsequent subtraction from the obtained results the systematic error calculated according to the model where $\mu_{rel}$ in the absorber plates is set to the relative permeability of air. According to this procedure the resulting axial force at $B_0$ of 3.81 T is equal to −19.15 kN on the "positive" endcap and +19.08 kN on the "negative" endcap. These values are 2.79 % larger than those reported in Ref. [3]. Those forces $F_z$, expected for the $\mu_{rel}$ value of 1.000588, were obtained from the forces $F_{z\,1.05}$ calculated for the $\mu_{rel}$ value of 1.05 and approximated then to the forces $F_z$ using an expression as follows:

$$F_z = 20 \cdot F_{z\,1.05}(\mu_{rel} - 1). \tag{2}$$

Calculating the model with 44 absorber plates inserted inside the superconducting coil requires about 23.5 hours of CPU time for any value of the relative permeability of absorbers. That is about 80.8 % larger than time needed to calculate the regular model [4] and is caused by the increasing the number of the spatial nodes in the model with absorbers to 12,858,342 that is 46.8 % larger than in the CMS magnet model without the stainless steel plates.

## 2 Calculating the magnetic forces for the orthogonal and angular misalignments

To estimate the forces to the HGCal hadronic compartments misaligned with respect to the coil axis, we have displaced and rotate the coil with respect to the $Z$ axis of the coordinate system in the vertical plane. Since the HGCal endcaps are strictly tied to the reference coordinate system, lowering the CMS coil by 10 mm along $Y$ axis as shown in Figure 4 is equivalent to a displacement of both "positive" and "negative" endcaps of the HGCal by 10 mm up with respect to the coil axis. We will call this displacement as Misalignment 1.

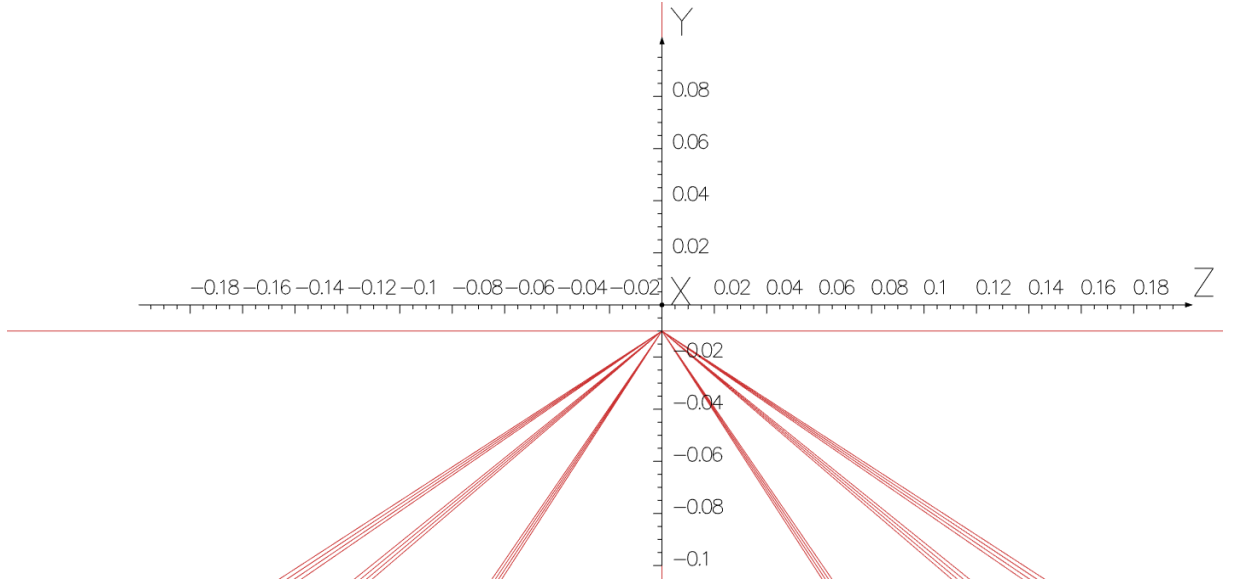

**Figure 4.** The coil axis (horizontal red line) lowered down by 10 mm along the $Y$ axis with respect to an origin of the coordinate system where the $Z$ axis coincides with the axes of the HGCal endcaps. The red triangles indicate the current direction in the coil central module far side at the positive $X$ coordinates.

Then, lowering the CMS coil by 5 mm along *Y* axis as shown in Figure 5 and rotating the coil by 0.05484° or 0.9571 mrad in a direction from *Z* to *Y* axis is equivalent to a rotation of the HGCal "positive" endcap outwards the coil axis in the point located at $X = Y = 0$ and $Z = 5.22391$ m, where the coil axis will cross the *Z* axis (Misalignment 2). The "negative" end cap in this case will be lifted with respect to the coil axis by 10 mm in the point located at $X = Y = 0$ and $Z = -5.22391$ m and then will be rotated by the same angle towards the coil axis (Misalignment 3).

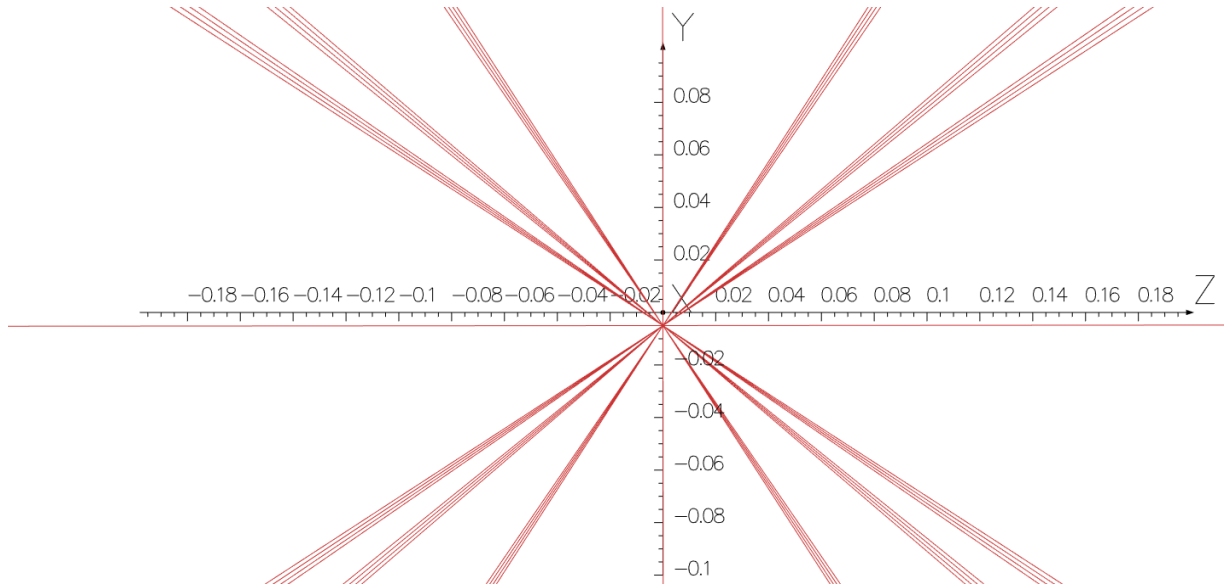

**Figure 5.** The coil axis (a red line rotated by 0.05484° or 0.9571 mrad with respect to the *Z* axis) lowered down by 5 mm along the *Y* axis with respect to an origin of the coordinate system where the *Z* axis coincides with the axes of the HGCal endcaps. The red triangles indicate the current direction in the coil central module far side (below the coil axis) at the positive *X* coordinates, and near side (above the coil axis) at the negative *X* coordinates.

In Figure 6, the "positive" endcap is displayed in accordance with Misalignment 1. In this case the vertical force $F_y$ is directed from the coil axis towards the endcap axis and is equal to 25.78 N, i. e. 2.6 N per 1 mm of displacement, that is vanished.

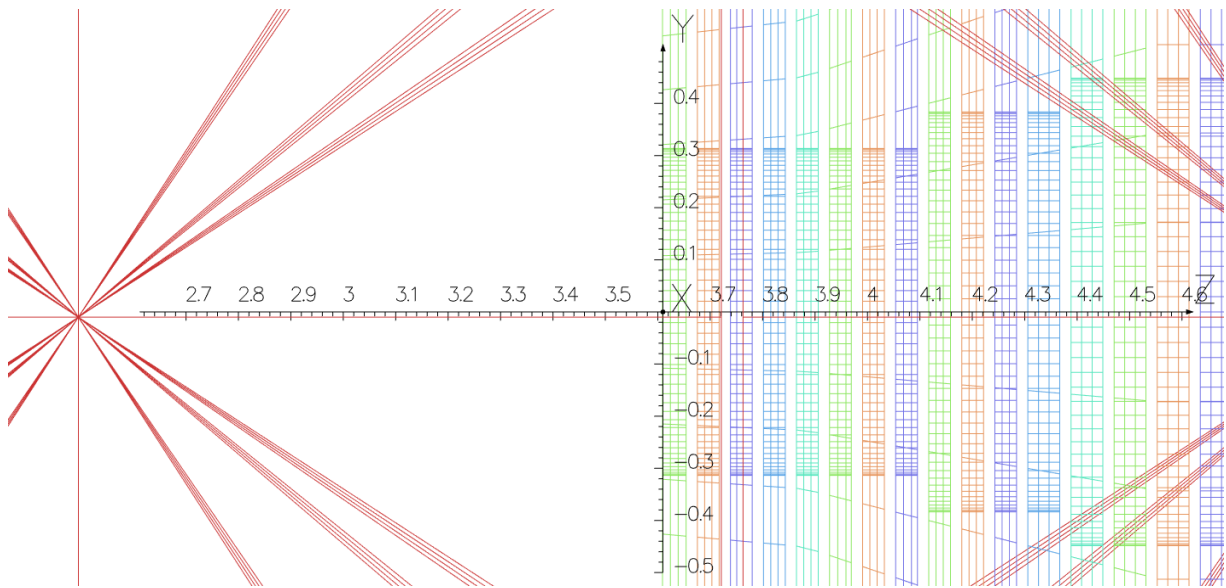

**Figure 6.** The "positive" HGCal endcap lifted by 10 mm with respect to the coil axis (Misalignment 1). The inner bores with radii of 0.3136 m (in the disks with numbers from 1 to 8), 0.3836 m (in the disks with numbers from 9 to 12), and 0.4448 m (in the disks with numbers from 13 to 16) are visible.

The torque $T_x$ with respect to the point located at $X = Y = 0$ and $Z = 5.22391$ m is equal to 3.35 N·m and is increasing the angle between the coil axis and the endcap axis. By dividing $T_x$ to $F_y$ we can get the $Z$ coordinate of the force application point equal to 5.098 m, i. e. 0.126 m from the torque application point.

In Figure 7, the "negative" endcap is displayed in accordance with Misalignment 1. In this case the vertical force $F_y$ is also directed from the coil axis towards the endcap axis and is equal to 25.97 N, i. e. 2.6 N per 1 mm of displacement, that is vanished.

The torque $T_x$ with respect to the point located at $X = Y = 0$ and $Z = -5.22391$ m is equal to $-3.87$ N·m and is also increasing the angle between the coil axis and the endcap axis. By dividing $T_x$ to $F_y$ we can get the $Z$ coordinate of the force application point equal to $-5.075$ m, i. e. 0.149 m from the torque application point.

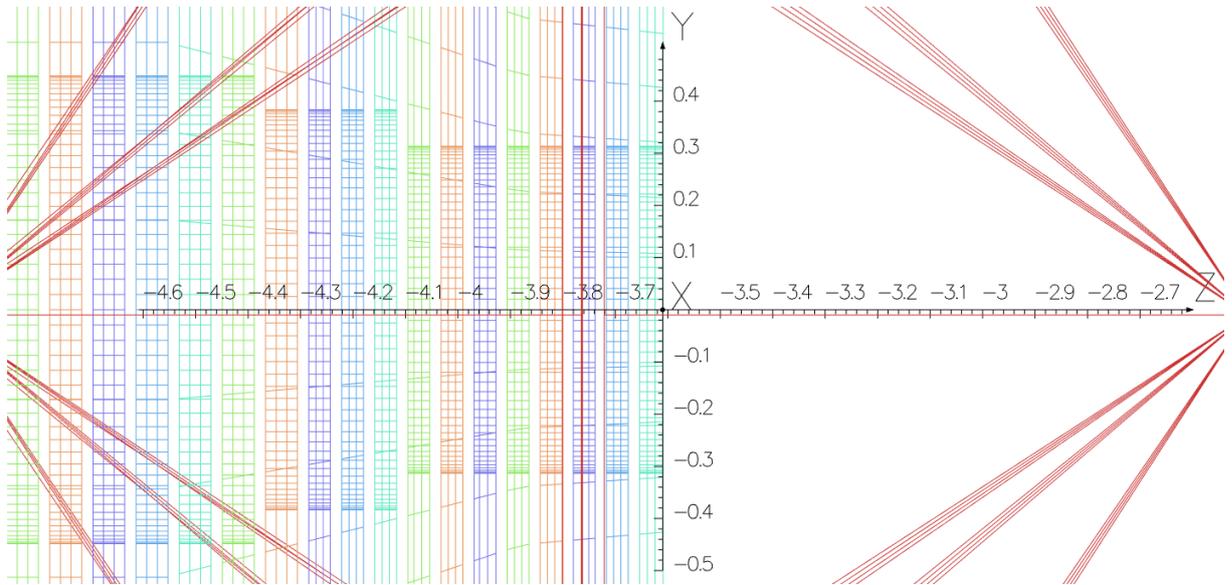

**Figure 7.** The "negative" HGCal endcap lifted by 10 mm with respect to the coil axis (Misalignment 1). The inner bores with radii of 0.3136 m (in the disks with numbers from −1 to −8), 0.3836 m (in the disks with numbers from −9 to −12), and 0.4448 m (in the disks with numbers from −13 to −18) are visible.

In Figure 8, the "positive" endcap is displayed in accordance with Misalignment 2.

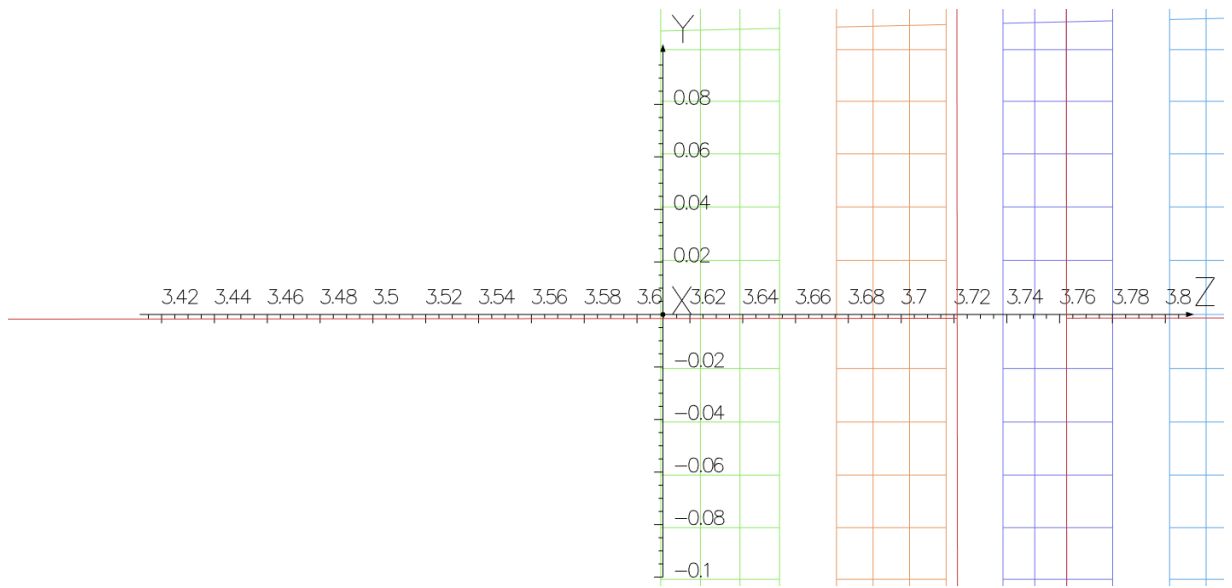

**Figure 8.** The "positive" HGCal endcap rotated by 0.05484° or 0.9571 mrad with respect to the coil axis in the point located at $X = Y = 0$ and $Z = 5.22391$ m (Misalignment 2).

In this case the vertical force $F_y$ is directed from the endcap axis towards the coil axis and is equal to $-24.47$ N, i. e. $-25.56$ N per 1 mrad of rotation, that is vanished.

The torque $T_x$ with respect to the point located at $X = Y = 0$ and $Z = 5.22391$ m is equal to $-48.84$ N·m and is decreasing the angle between the coil axis and the endcap axis. By dividing $T_x$ to $F_y$ we can get the $Z$ coordinate of the force application point equal to 3.228 m, i. e. 1.996 m from the torque application point.

In Figure 9, the "negative" endcap is displayed in accordance with Misalignment 3. In this case the vertical force $F_y$ is directed from the coil axis towards the endcap axis and is equal to 49.95 N, i. e. 5.0 N per 1 mm of displacement, that is vanished.

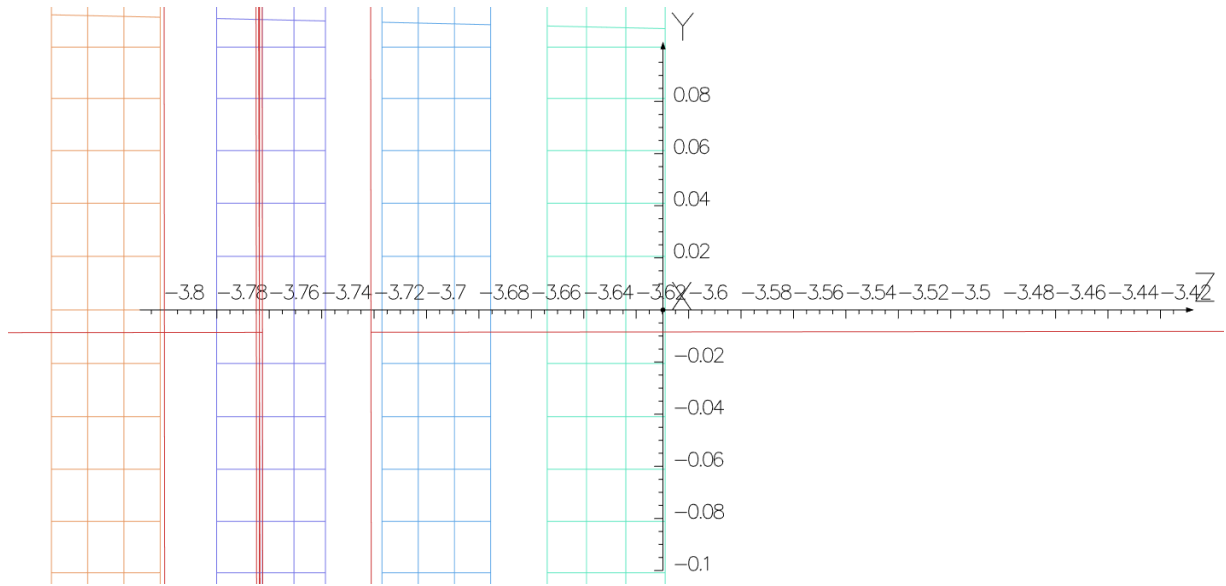

**Figure 9.** The "negative" HGCal endcap lifted by 10 mm with respect to the coil axis and rotated by 0.05484° or 0.9571 mrad in the point located at X = Y = 0 and Z = 5.22391 m towards the coil axis (Misalignment 3).

The torque $T_x$ with respect to the point located at $X = Y = 0$ and $Z = -5.22391$ m is equal to $-53.26$ N·m and is increasing the angle between the coil axis and the endcap axis. By dividing $T_x$ to $F_y$ we can get the $Z$ coordinate of the force application point equal to $-4.158$ m, i. e. 1.066 m from the torque application point.

## 3 Conclusions

In this study, the calculation of the electromagnetic forces acting to the stainless steel hadronic compartments of a new high granularity calorimeter is performed using the modified three-dimensional model of the Compact Muon Solenoid (CMS) magnet. With a relative permeability value of 1.000588 the maximum axial force acting to the 22 plates in the hadronic compartment at the CMS central magnetic flux density of 3.81 T is 19.15 kN. The side forces to the endcap compartment due to displacements and rotations of the endcap axis with respect to the CMS coil axis are vanished and do not exceed 5 N per 1 mm of displacement or 26 N per 1 mrad of rotation.

## References

[1] CMS Collaboration. The CMS experiment at the CERN LHC. *J. Instrum.* **2008**, *3*, S08004. https://doi.org/10.1088/1748-0221/3/08/S08004.
[2] CMS Collaboration. The Phase-2 Upgrade of the CMS endcap calorimeter. Technical Design Report. *CERN-LHCC-2017-023, CMS-TDR-019*; CERN: Geneva, Switzerland. **2018**, pp. 11–20. ISBN: 978-92-9083-459-5. Available online: https://cds.cern.ch/record/2293646 (accessed on 1 October 2023).
[3] Klyukhin, V. on behalf of the CMS Collaboration. Calculation of Forces to the High Granularity Calorimeter Stainless Steel Absorber Plates in the CMS Magnetic Field. *Symmetry*, **2023** *15*, 2017. https://doi.org/10.3390/sym15112017.
[4] Klyukhin, V. Design and Description of the CMS Magnetic System Model. *Symmetry* **2021**, *13*, 1052. https://doi.org/10.3390/sym13061052.
[5] *TOSCA/OPERA-3d 18R2 Reference Manual*; Cobham CTS Ltd.: Kidlington, UK, 2018; pp. 1–916.
[6] Simkin, J.; Trowbridge, C. Three-dimensional nonlinear electromagnetic field computations, using scalar potentials. In *IEE Proceedings B Electric Power Applications*; Institution of Engineering and Technology (IET): London, UK, **1980**; *127*, 368–374. https://doi.org/10.1049/ip-b.1980.0052.
[7] Tamm, I.E. Fundamentals of the theory of electricity; Ninth Russian Edition, Moscow: Nauka, 1976; pp. 385–387 (in Russian).